\documentclass[pre,twocolumn,superscriptaddress,nofootinbib]{revtex4-2}
\usepackage{graphicx,chemarr,color,hyperref}
\usepackage[version=4]{mhchem}

\usepackage{amssymb,amsmath,textgreek}

\begin{document}

\title{Near-critical gene expression in embryonic boundary precision}

\author{Michael Vennettilli}
\affiliation{Department of Physics and Astronomy, University of Pittsburgh, Pittsburgh, Pennsylvania 15260, USA}
\affiliation{AMOLF, Science Park 104, 1098 XG, Amsterdam, The Netherlands}

\author{Krishna P.\ Ramachandran}
\affiliation{Department of Physics and Astronomy, University of Pittsburgh, Pittsburgh, Pennsylvania 15260, USA}

\author{Andrew Mugler}
\email{andrew.mugler@pitt.edu}
\affiliation{Department of Physics and Astronomy, University of Pittsburgh, Pittsburgh, Pennsylvania 15260, USA}

\begin{abstract}
Embryonic development relies on the formation of sharp, precise gene expression boundaries. In the fruit fly {\it Drosophila melanogaster}, boundary formation has been proposed to occur at a dynamical critical point. Yet, in the paradigmatic case of the {\it hunchback} ({\it hb}) gene, evidence suggests that boundary formation occurs in a bistable regime, not at the dynamical critical point. We develop a minimal model for {\it hb} expression and identify a single parameter that tunes the system from its monostable regime to its bistable regime, crossing the critical point in between. We find that boundary precision is maximized when the system is weakly bistable---near, but not at, the critical point---optimally negotiating the tradeoff between two key effects of bistability: sharpening the boundary and amplifying its noise.  Incorporating the diffusion of Hb proteins into our model, we show that boundary precision is maximized simultaneously at an optimal degree of bistability and an optimal diffusion strength. Our work elucidates design principles of precise boundary formation and has general implications for pattern formation in multicellular systems.
\end{abstract}

\maketitle

\section{Introduction}

In a developing fruit fly embryo, gene expression profiles provide the body plan. The so-called gap genes form a series of profiles that shift from low to high expression (or vice versa) at various positions along the length of the embryo \cite{gilbert2023developmental}. These positions ultimately delineate the segments of the fly's body. Therefore, it is important that a boundary between low and high expression levels is sharply defined and has minimal variation from embryo to embryo. In the the fruit fly {\it Drosophila melanogaster}, boundary positions are specified with a precision of 1\% of the embryo length, equivalent to the distance between individual cell nuclei \cite{petkova2019optimal}.

\begin{figure}[b]
\includegraphics[width=1\columnwidth]{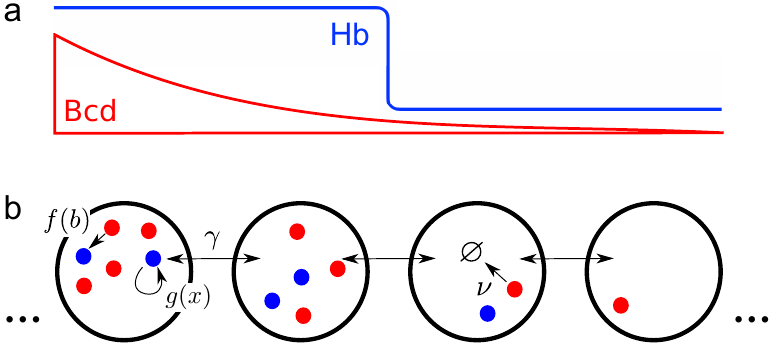}
\caption{Hunchback (Hb) boundary formation. (a) Schematic of sharp Hb profile and exponential Bicoid (Bcd) profile along the anterior-posterior axis. (b) Hb is activated by both Bcd and itself, diffuses, and degrades.}
\label{cartoon}
\end{figure}

A well-studied gap gene is {\it hunchback} ({\it hb}), which forms a sharp boundary between high expression in the anterior half of the embryo and low expression in the posterior half (Fig.\ \ref{cartoon}a). Hb is activated by the Bicoid (Bcd) protein \cite{struhl1989gradient, driever1989bicoid}, which itself forms an approximately exponential profile along the anterior-posterior axis soon after fertilization \cite{houchmandzadeh2002establishment, gregor2007stability} (Fig.\ \ref{cartoon}a). Therefore, much of the sharpness of the Hb boundary is attributed to cooperative activation by Bcd \cite{struhl1989gradient, driever1989bicoid, ma1996drosophila, crauk2005bicoid, gregor2007probing}. However, Hb also activates itself \cite{treisman1989products, simpson1994synergy}. Constructs without this positive feedback form Hb boundaries that are less sharp than in wild-type embryos \cite{lopes2008spatial}, raising the possibility that positive feedback sharpens the boundary.

Positive feedback can lead to a bifurcation from monostable to bistable dynamics \cite{strogatz2018nonlinear}. The bifurcation point has the usual properties of a critical point, including critical slowing down, strong fluctuations, and---when extended over space, as in an embryo---long-range correlations \cite{krotov2014morphogenesis, vennettilli2020multicellular}. Biological systems have been conjectured to sit at such critical points in their parameter space \cite{mora2011biological, munoz2018colloquium}. In fact, all of the above properties of criticality have been observed in {\it Drosophila} in the context of mutually repressive gene pairs whose boundaries cross, leading to the proposal that these boundaries are tuned to a critical point \cite{krotov2014morphogenesis}. Because mutual repression is a net positive feedback, this suggests that Hb self-activation, also a positive feedback, may be tuned to the critical point between its monostable and bistable dynamics.

On the other hand, from the perspective of boundary sharpness,  bistable rather than critical dynamics could be advantageous: weakly separated expression levels on either side of the boundary could be driven by the bistability to more strongly separated high and low stable states, respectively. Indeed, computational work has suggested that bistability is necessary to explain the sharpness of the Hb boundary, as well as to explain the experimentally observed effects of removing Hb self-activation and removing the Bcd protein \cite{lopes2008spatial}. However, in addition to being sharp, it is important that a boundary have minimal positional variation, or ``noise'' \cite{gregor2007probing}. Bistability may introduce large positional variation, as expression levels at positions near the boundary are susceptible to large flips between the high and low states. The above computational work considered only boundary sharpness, not boundary noise.

Additionally, the Hb profile forms before cells form their outer membranes, meaning that Hb proteins could potentially diffuse throughout the embryo \cite{gregor2007probing}. Previous theoretical work has shown that diffusion reduces the boundary sharpness, but also reduces the boundary noise, leading to an optimal diffusion coefficient \cite{erdmann2009role}. This raises the question of whether Hb feedback and Hb diffusion could synergistically improve boundary formation. This previous study did not consider the effects of feedback.

Here, we introduce a model of Hb profile formation that includes activation by Bcd, self-activating feedback, and diffusion. By mapping the model to a canonical form for this type of bifurcation, we are able to characterize the system entirely in terms of two parameters, one describing the strength of the feedback (and the associated distance to the critical point), and the other describing the strength of diffusion. We capture both boundary sharpness and boundary noise in a single measure, and find that this measure is simultaneously optimized at a particular feedback strength and diffusion strength. The optimal feedback places the boundary weakly in the bistable regime, near but not at the critical point, which negotiates the tradeoff between the two anticipated effects of bistability: sharpening the boundary and amplifying its noise. Our work elucidates design principles of precise boundary formation, and we discuss its implications for embryonic development in particular and pattern formation in general.

\section{Results}

\subsection{Model}

Because the Bcd and Hb profiles vary only along the anterior-posterior axis of the embryo, it is sufficient to consider a line of $N$ cell nuclei along this axis (Fig.\ \ref{cartoon}b). We take $N=40$, which is less than the full embryo length but sufficiently large to focus on the properties of the boundary region, including the effects of Hb diffusion.

The Bcd molecule number in nucleus $i$ is given by the exponential function
\begin{equation}
\label{bcd}
b_i = b_*e^{-(i-i_*)\ell/\lambda},
\end{equation}
where $i_*=N/2$ is the position of the Hb boundary, and experimental estimates set the nuclear spacing $\ell = 8.5$ $\mu$m \cite{gregor2007probing}, exponential lengthscale $\lambda = 120$ $\mu$m \cite{houchmandzadeh2002establishment, gregor2007probing, erdmann2009role}, and molecule number at the boundary $b_* = 690$ \cite{gregor2007probing}.

The Hb molecule number in nucleus $i$ obeys the dynamics
\begin{equation}
\label{hb}
\frac{dx_i}{dt} = f(b_i) + g(x_i) + \gamma \Delta_ix_i - \nu x_i,
\end{equation}
where
\begin{align}
\label{activation}
f(b_i) &= k_b\frac{b_i^B}{b_i^B + b_{1/2}^B},\\
g(x_i) &= k_x\frac{x_i^H}{x_i^H + x_{1/2}^H},\\
\gamma \Delta_ix_i &= \gamma(x_{i+1} + x_{i-1} - 2x_i),
\end{align}
and $\nu x_i$ account for activation by Bcd, self-activation, diffusion, and degradation, respectively (Fig.\ \ref{cartoon}b).
In Eq.\ \ref{activation}, we set $b_{1/2}=b_*$ to ensure that the activation is half-maximal at the boundary, where $b_i=b_*$ (Eq.\ \ref{bcd}).  Peak Hb molecule numbers are estimated to be at least $820$$-$$1300$ per nucleus \cite{zamparo2009statistical}, and therefore we set $x_{1/2} = 820$ at the lower end of that range. This ensures that Hb molecule numbers are near that value at the boundary, and higher but on that order at the anterior peak. We set the Hill coefficients to the number of Bcd and Hb binding sites on the Hb promoter that are sufficient to produce the wild-type Hb profile in experiments, namely $B = 3$ \cite{driever1989bicoid} and $H = 2$ \cite{lopes2008spatial}, respectively. Previous modeling studies have assumed a degradation rate of $\nu \sim 1$ h$^{-1}$ or more \cite{erdmann2009role, jaeger2007known}, and we later find that above this value our results are unaffected. The remaining parameters $k_b$, $k_x$, and $\gamma$ are either varied or determined by the transformation of Eq.\ \ref{hb} to a canonical form, as described in the next section. All parameters are summarized in Table \ref{params}.

\begin{table}[b]
\begin{center}
\begin{tabular}{|l|l|l|l|}
\hline
$\ell$ & Nuclear spacing & 8.5 $\mu$m & \cite{gregor2007probing} \\
\hline
$\lambda$ & Bcd profile lengthscale & 120 $\mu$m & \cite{houchmandzadeh2002establishment, gregor2007probing, erdmann2009role} \\
\hline
$b_*$ & Bcd number at boundary & 690 & \cite{gregor2007probing} \\
\hline
$x_{1/2}$ & Feedback half-maximal Hb no. & 820 & \cite{zamparo2009statistical} \\
\hline
$B$ & Activation Hill coefficient & 3 & \cite{driever1989bicoid} \\
\hline
$H$ & Feedback Hill coefficient & 2 & \cite{lopes2008spatial} \\
\hline
$\nu$ & Hb degradation rate & $\gtrsim 1$ h$^{-1}$ & \cite{erdmann2009role, jaeger2007known} \\
\hline
$k_b$ & Activation strength & Eq.\ \ref{kb} & \\
\hline
$k_x$ & Feedback strength & Eq.\ \ref{kx} & \\
\hline
$\gamma$ & Diffusion rate & varied & \\
\hline
\end{tabular}
\caption{Parameters used in this work.}
\label{params}
\end{center}
\end{table}

Because the time for a Bcd molecule to diffuse between nuclei is much smaller than the time between bindings to the Hb promoter, diffusive averaging allows us to neglect Bcd molecule number fluctuations in Eq.\ \ref{bcd} \cite{erdmann2009role}. However, Hb molecule number fluctuations in Eq.\ \ref{hb} will affect the boundary noise and will therefore be considered later when we describe our numerical results from stochastic simulations.

\subsection{Canonical form}
By expanding the right-hand side of Eq.\ \ref{hb} up to third order in $x_i$, it can be placed in the normal form of an imperfect pitchfork bifurcation \cite{strogatz2018nonlinear},
\begin{equation}
\label{ising}
\frac{dm_i}{d\tau} = h_i - \theta m_i - \frac{1}{3}m_i^3 + \Gamma\Delta_im_i.
\end{equation}
Specifically, defining
\begin{equation}
m_i \equiv \frac{x_i-x_c}{x_c}
\end{equation}
shifts and scales $x_i$ to remove the quadratic term from Eq.\ \ref{ising}, where
\begin{equation}
\label{xc}
x_c \equiv \left(\frac{H-1}{H+1}\right)^{1/H}x_{1/2}
\end{equation}
is the value of $x_i$ at which the second derivative of the the right-hand side of Eq.\ \ref{hb} with respect to $x_i$ vanishes. Meanwhile, scaling time as
\begin{equation}
\tau \equiv \frac{(H^2-1)^2}{16Hx_c}k_xt
\end{equation}
leaves the cubic term in Eq.\ \ref{ising} with the normal coefficient of $-1/3$. The remaining expansion coefficients are then
\begin{align}
\label{h}
h_i &\equiv \frac{16Hf(b_i) + 8(H-1) k_x - 16H\nu x_c}{(H^2-1)^2 k_x}, \\
\label{theta}
\theta &\equiv \frac{16H\nu x_c - 4(H^2-1)k_x}{(H^2-1)^2 k_x}, \\
\Gamma &\equiv \frac{16Hx_c}{(H^2-1)^2 k_x}\gamma,
\end{align}
where $\Gamma$ is the time-rescaled diffusion rate.

From a statistical mechanics perspective, Eq.\ \ref{ising} has the same form as the mean-field Ising model, where $m_i$ is the magnetization, $h_i$ is the dimensionless magnetic field, and $\theta$ is the reduced temperature \cite{byrd2019critical, erez2019universality, erez2020cell, vennettilli2020multicellular}. In this context, $m_i$ is a position-dependent order parameter indicating whether the Hb molecule number is above ($m_i>0$) or below ($m_i<0$) the characteristic value $x_c$, and $h_i$ biases that number higher ($h_i>0$) or lower ($h_i<0$). The Ising transition from a disordered state ($\theta > 0$) to a symmetry-breaking ordered state ($\theta < 0$) corresponds in our system to the transition from monostability to bistability. Note that the activation by Bcd $f(b_i)$ enters only in the ``external'' field $h_i$, whereas the ``internal'' temperature parameter $\theta$ depends only on the auto-activating feedback.

The critical point occurs at $h_i = \theta = 0$. It is natural to set $h_i = 0$ at the boundary ($i=i_*$) because then this field pushes Hb molecule numbers higher and lower on either side. Applying this condition to Eq.\ \ref{h}, recognizing that $f(b_*) = k_b/2$ from Eq.\ \ref{activation}, and solving for $k_b$ gives $k_b = 2\nu x_c - (H-1)k_x/H$. Solving Eq.\ \ref{theta} for $k_x$ and inserting it into this expression gives
\begin{align}
\label{kx}
k_x &= \frac{16H\nu x_c}{(H^2-1)[(H^2-1)\theta+4]}, \\
\label{kb}
k_b &= 2\nu x_c \left\{ 1 - \frac{8}{(H+1)[(H^2-1)\theta + 4]} \right\}.
\end{align}
Recalling Eq.\ \ref{xc}, we see that Eqs.\ \ref{kx} and \ref{kb} give the feedback and activation strengths entirely in terms of the reduced temperature $\theta$ and known parameters (Table \ref{params}). We thus have a single parameter $\theta$ that determines the distance to the critical point: criticality corresponds to $\theta = 0$, whereas the monostable and bistable regimes correspond to $\theta > 0$ and $\theta < 0$, respectively. As we tune $\theta$, $k_x$ and $k_b$ follow accordingly from Eqs.\ \ref{kx} and \ref{kb}.

We note that $\theta$ can be increased, but not decreased, without bound. The limit $\theta\to\infty$ corresponds to no feedback ($k_x\to0$ by Eq.\ \ref{kx}). From below, $\theta$ is limited by the requirements $k_x\ge0$ and $k_b\ge0$. Using Eqs.\ \ref{kx} and \ref{kb}, these requirements imply $\theta\ge-4/(H^2-1)$ and $\theta\ge-4/(H+1)^2$, respectively, with the latter being the sharper bound. In particular, with $H=2$ (Table \ref{params}), we have $\theta\ge-4/9$, which defines the most strongly bistable regime.

\subsection{Numerical results}

We perform stochastic simulations \cite{gillespie1977exact} of the Hb molecular reactions in Eq.\ \ref{hb} (see \cite{code} for the code). Specifically, in nucleus $i$, a Hb molecule is produced with propensity $f(b_i)+g(x_i)$, degrades with propensity $\nu x_i$, and diffuses to a neighboring nucleus with propensity $\gamma x_i$. All parameters are as in Table \ref{params} (with $\nu = 5$ h$^{-1}$), and we vary the feedback parameter $\theta$ and the diffusion parameter $\gamma$.

Hb is deposited maternally and is already present shortly after fertilization, preferentially in the anterior half of the embryo \cite{tautz1988regulation, struhl1992control}, suggesting that we initialize with $x_i$ a decreasing function of $i$. We use
\begin{equation}
\label{init}
x_i(0) = A\left\{1-\tanh\left[\frac{(i-i_*)\ell}{\lambda}\right]\right\},
\end{equation}
and we later find that for $A\gtrsim x_c$ our results are unaffected; we use $A = x_c$ here. The Hb boundary is formed during cell cycle 14, which lasts at least 65 minutes \cite{foe1983studies}, and therefore we simulate for $T=100$ minutes.

\begin{figure}
\includegraphics[width=1\columnwidth]{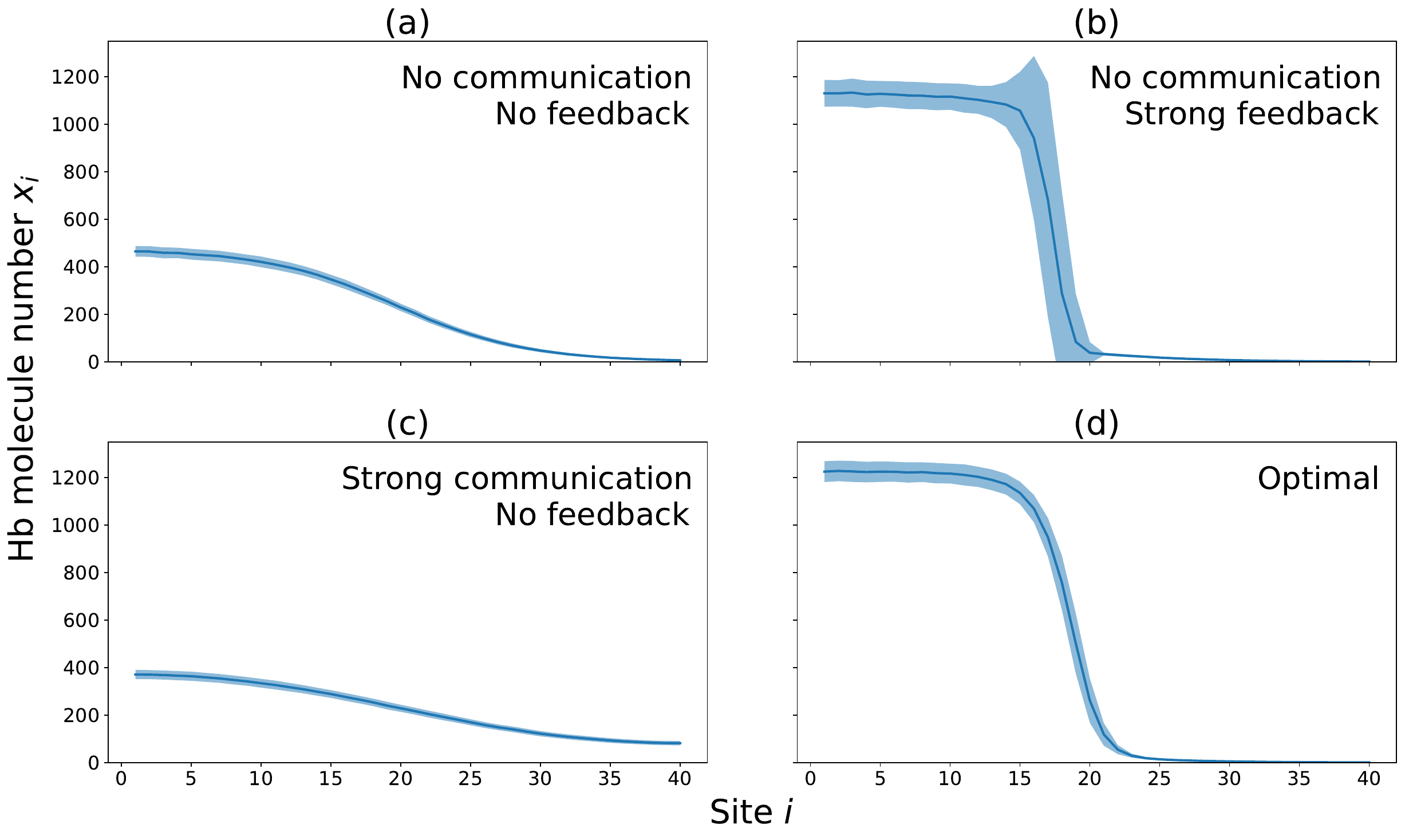}
\caption{Hb profiles from stochastic simulations (means and standard deviations across $10^5$ trials). (a) No feedback ($k_x = 0$), no diffusion ($\gamma=0$). (b) Strong feedback ($\theta = -0.3$), no diffusion. (c) No feedback, strong diffusion ($\gamma/\nu = 200$). (d) Optimal feedback and diffusion from Fig.\ \ref{heatmap}. Parameters are as in Table \ref{params} with $\nu = 5$ h$^{-1}$ and $A = x_c = 473$ (Eq.\ \ref{init}).}
\label{profiles}
\end{figure}

Without feedback ($\theta\to\infty$) or diffusion ($\gamma=0$), we find that Hb forms a moderately sharp boundary due to cooperative activation by Bcd alone (Fig.\ \ref{profiles}a). Without diffusion, but with sufficiently strong feedback to induce bistability ($\theta<0$), the boundary is sharpened significantly (Fig.\ \ref{profiles}b), consistent with previous work \cite{lopes2008spatial}. At the same time, however, we see that the embryo-to-embryo variability is significantly increased (envelope in Fig.\ \ref{profiles}b) due to the large switching noise associated with bistability. Alternatively, without feedback, but with strong diffusion, the variability is reduced, but the boundary is more shallow (Fig.\ \ref{profiles}c), also consistent with previous work \cite{erdmann2009role}.

To capture boundary sharpness and variability in a single quantity \cite{erdmann2009role}, we introduce the dimensionless sensitivity measure
\begin{equation}
\label{S}
S = \frac{\bar{x}_{i_*-1}-\bar{x}_{i_*}}{(\sigma_{i_*-1}+\sigma_{i_*})/2}.
\end{equation}
This measure increases with sharpness because the numerator is the difference between mean Hb numbers in neighboring nuclei at the boundary. It decreases with variability because the denominator is the average of the standard deviations of Hb numbers in these cells across simulation trials.

\begin{figure}
\includegraphics[width=1\columnwidth]{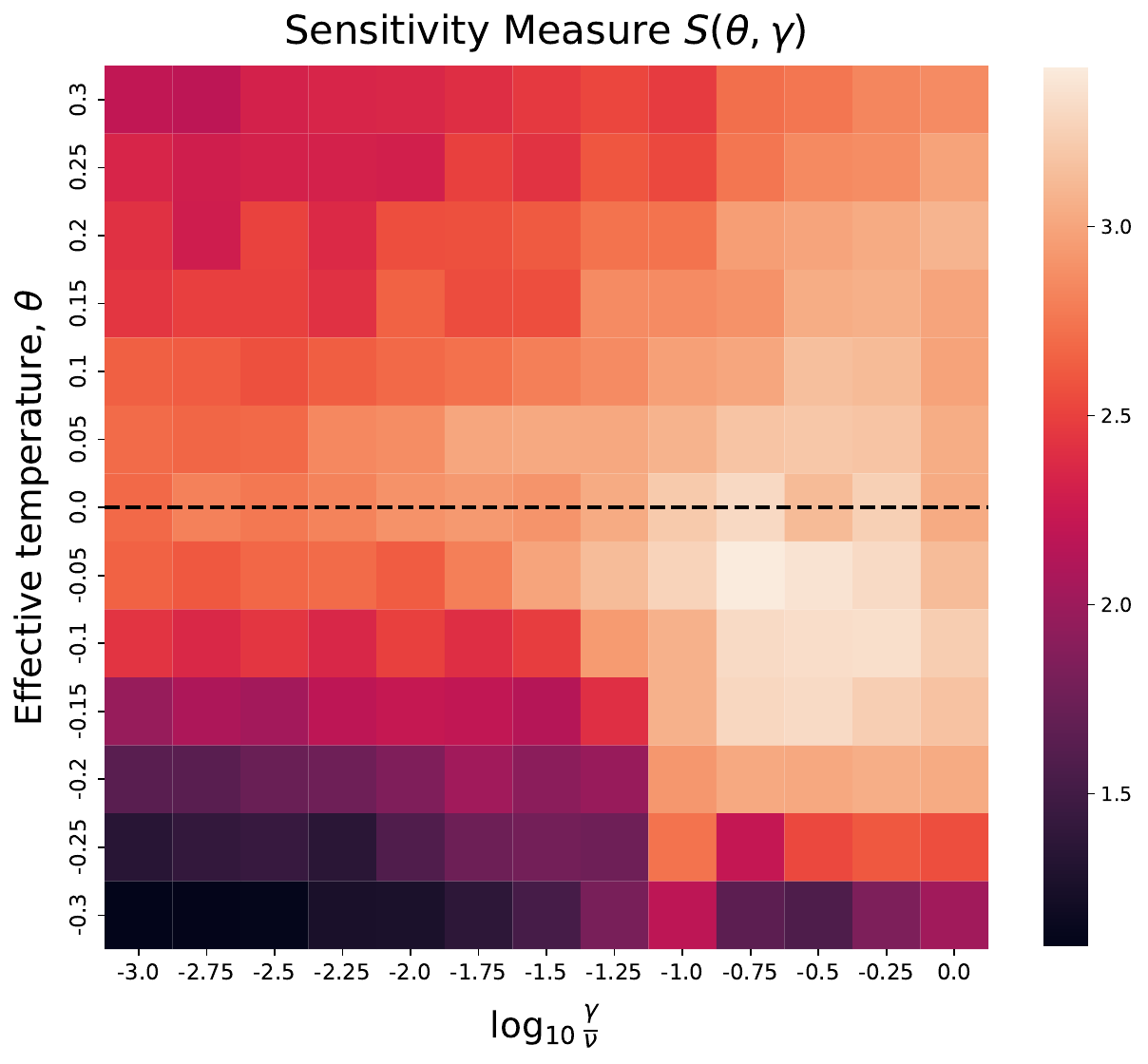}
\caption{Boundary sensitivity (Eq.\ \ref{S}) from simulations exhibits an optimum as a function of feedback $\theta$ and diffusion $\gamma/\nu$. Dashed line indicates critical point $\theta = 0$. Parameters are as in Table \ref{params} with $\nu = 5$ h$^{-1}$ and $A = x_c = 473$ (Eqs.\ \ref{init} and \ref{xc}).}
\label{heatmap}
\end{figure}

To investigate the simultaneous effects of feedback and diffusion, we plot $S$ against $\theta$ and $\gamma$ in Fig.\ \ref{heatmap}. We see that boundary formation is optimally sharp and precise (maximum $S$) when $\theta \approx -0.1$ and $\gamma/\nu\approx0.3$. The optimal $\theta$ value indicates a response that is near, but not at, the critical point ($\theta = 0$) and weakly in the bimodal regime (about a quarter of the way to the maximally bistable value of $\theta = -4/9$). This value optimally negotiates the tradeoff illustrated in Fig.\ \ref{profiles}b: bistability sharpens the boundary but increases noise. The optimal $\gamma/\nu$ value indicates that the diffusion rate should be on the same order as the degradation rate. This makes sense because the latter sets the natural timescale in the system. Equivalently, since the diffusion coefficient is $D \sim \gamma \ell^2$, it implies that the diffusion lengthscale $\sqrt{D/\nu}$ should be on the order of the nuclear spacing $\ell$. Together, the optimal $\theta$ and $\gamma/\nu$ values result in a boundary that is both sharp and precise, as illustrated in Fig.\ \ref{profiles}d.

\subsection{Robustness of near-critical optimum}

All parameters are either varied or known from experiments (Table \ref{params}) except two: the initial anterior Hb molecule number $A$ and the Hb degradation rate $\nu$. Here we ask whether our main finding, that near-critical feedback is optimal ($\theta^* \approx -0.1$), is sensitive to our choices for $A$ and $\nu$.

\begin{figure}
\includegraphics[width=1\columnwidth]{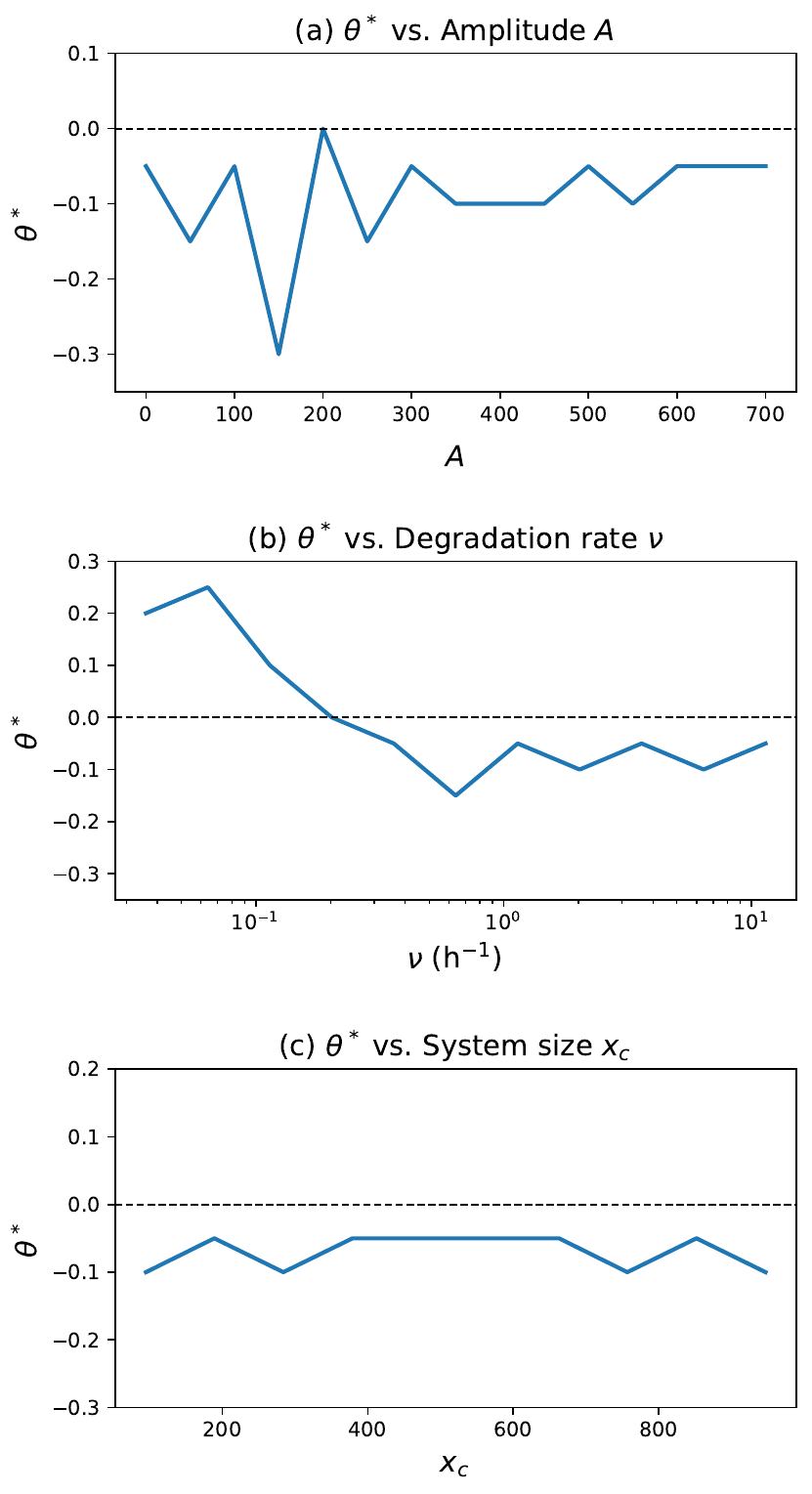}
\caption{Dependence of the optimum $\theta^*$ on the (a) initial Hb molecule number in the anterior $A$, (b) the Hb degradation rate $\nu$, and (c) the characteristic Hb molecule number $x_c$. Except when varied, parameters are as in Table \ref{params} with $\nu = 5$ h$^{-1}$, $A = 473$, and $\gamma/\nu$ at its optimal value in Fig.\ \ref{heatmap}. When $\nu$ is varied we keep $\gamma/\nu$ at this value. $x_c$ is varied by changing $x_{1/2}$ (Eq.\ \ref{xc}).}
\label{robust}
\end{figure}

We see in Fig.\ \ref{robust}a that $\theta^*$ is variable at small $A$ but then unaffected by $A$ beyond $A \approx x_c = 473$. The variability at small $A$ is due to metastable trapping: when $A$ is small, the Hb molecule number in the anterior is initialized in the low state, even though the large Bcd molecule number there makes the high state more stable. The Hb molecule number can get trapped in the low state, and, in some trials, noise may not be enough to kick it to the more stable high state. The fact that the transition occurs at $A \approx x_c$ makes sense because $x_c$ sets the barrier between the low- and high-molecule-number states in the bistable regime. Thus, our model predicts that the anterior deposit of maternal Hb \cite{tautz1988regulation, struhl1992control} should be on the order of its ultimate level at the boundary or larger, in order to avoid metastable trapping.

We see in Fig.\ \ref{robust}b that $\theta^*$ is unaffected by $\nu$ beyond $\nu \approx 1/T = 0.6$ h$^{-1}$. This makes sense because $1/\nu$, which sets the response timescale, should be smaller than $T$, the total simulation time, in order for the system to reach steady state. Although we are unaware of measurements of the Hb degradation rate, most modeling studies use rates on the order of or much faster than once per hour \cite{erdmann2009role, jaeger2007known}, which is where we find that $\theta^*$ is unaffected by $\nu$. It has also been suggested that Hb mRNA is actively degraded \cite{desponds2020mechanism}, and once per hour is a very modest rate for any active degradation.

Finally, we ask whether the fact that the optimal feedback is near, not at, the critical point is because of the finite size of the system. Does the optimum tend toward the critical point as the system size increases? In this model, the system size is given by the molecule number scale $x_c$: all of the critical exponents tend toward their expected values as $x_c \to \infty$ \cite{byrd2019critical, erez2019universality}. Therefore, we plot $\theta^*$ against $x_c$ in Fig.\ \ref{robust}c, focusing on the order of magnitude surrounding the experimental estimate ($x_c = 473$). We see that $\theta^*$ does not get closer to the critical point ($\theta=0$) as $x_c$ increases, but instead remains at $\theta^*\approx -0.1$. This result suggests that the benefit of weak bistability for boundary sensitivity is not a finite-size effect, but rather arises from the fundamental tradeoff between sharpening the boundary and reducing its noise.

\section{Discussion}

We have identified the basic physical requirements for forming a steady-state morphogen boundary that is both sharp and precise. By mapping the Bcd-Hb problem to a canonical thermodynamic form, we have reduced the problem to two control parameters: feedback strength and diffusion. The mapping has allowed us to ask whether optimal boundary formation occurs at a critical point. We find that it occurs near, but not at, the critical point, in the weakly bistable regime. This regime optimally negotiates the tradeoff between sharpening the boundary and reducing the variability in its position.

Previous work identified the benefit of bistability for sharpening the Hb boundary \cite{lopes2008spatial} but did not consider positional variability. Doing so, we find that the optimal bistability is weak, placing the system close to the critical point between its bistable and monostable regimes. Separate work identified experimental signatures that suggest the gap gene network sits at a critical point \cite{krotov2014morphogenesis}. Focusing on the Bcd-Hb regulation specifically, we find that sitting at its critical point would lose some of the boundary-sharpening benefit of bistability. Thus, according to our model, the optimal design is neither critical nor strongly bistable, but in between. It is an interesting open question whether some of the identified signatures of criticality \cite{krotov2014morphogenesis} would persist in this near-critical regime.

Previous work predicted, as we do here, an optimal Hb diffusion coefficient due to the tradeoff between diffusion reducing the boundary sharpness and reducing the boundary noise \cite{erdmann2009role}. Specifically, that work reported an optimal diffusion coefficient of $D\approx0.1$ $\mu$m$^2$/s when the degradation rate was $\nu = 0.4$ h$^{-1}$ (Fig.\ 2b of \cite{erdmann2009role}), corresponding to a rate ratio of $(D/\ell^2)/\nu\sim 10$. Here, we find an optimal rate ratio of $\gamma/\nu\approx0.3$. The difference in these values is likely due to the presence of Hb feedback in our model, which was not considered in the previous work. This comparison suggests that while the emergence of an optimal diffusion coefficient for boundary formation is generic, its value can be strongly affected by the presence of regulatory feedback.

Our work identifies near-critical bistability as a general design principle for reliable boundary formation. It would be interesting to explore whether this principle holds for other types of positive feedback such as mutual repression, and for larger networks such as the gap genes. More broadly, our findings may have implications for the formation of reliable boundaries in contexts beyond morphogenesis, such as cluster formation, Turing patterns, and wave propagation, all of which are common in systems from the subcellular to the multicellular scale.

\section*{Acknowledgments}

All authors were supported by the National Science Foundation (PHY-2118561).


%

\end{document}